\documentstyle[12pt]{article}

\newcommand {\beq}{\begin{eqnarray}}
\newcommand {\eeq}{\end{eqnarray}}
\newcommand {\beqs}{\begin{eqnarray*}}
\newcommand {\eeqs}{\end{eqnarray*}}

\newcommand {\bz}{\bf Z}
\newcommand {\br}{\bf R}

\newcommand {\hp}{hep-th/}
\newcommand{\be}{\begin{equation}}
\newcommand{\ee}{\end{equation}}
\newcommand{\ba}{\begin{eqnarray}}
\newcommand{\ea}{\end{eqnarray}}
\newcommand{\non}{\nonumber\\}
\newcommand{\Eq}[1]{(\ref{#1})}
\newcommand{\dsp}{\displaystyle}
\newcommand{\eqn}[1]{(\ref{#1})}

\begin{document}
\baselineskip=0.6cm

%%%%%%%%%%%%%%%%%%% titlepage %%%%%%%%%%%%%%%%%%%%%
\begin{titlepage}
\nopagebreak
%
%
%%%%%%%%%%%%%%%%%%%%%%%%%%%%%%%%%%%%%%%%%%%%%%%%%%%

\begin{flushright}
May 1996\hfill YITP/96-15\\
OU-HET 243\\
hep-th/9605023
\end{flushright}

\vfill
\begin{center}
{\LARGE BPS mass spectrum from D-brane action}
\vskip 20mm

{\large H.~Ishikawa${}^{\dag \S}$,
Y.~Matsuo${}^{\dag}$,
Y.~Sugawara${}^{\ddag}$,
K.~Sugiyama${}^{\dag \S}$
}
\vskip 10mm
${}^{\dag}$
{\sl Yukawa Institute for Theoretical Physics, Kyoto University} \\
{\sl Kitashirakawa, Kyoto 606-01, Japan}\\
\vskip 2mm
${}^{\ddag}$
{\sl Department of Physics, Osaka University}\\
{\sl Machikaneyama 1-1, Toyonaka, Osaka 560, Japan}\\
\end{center}
\vfill

\begin{abstract}
We derive the BPS mass formulae of the Dirichlet branes
from the Born-Infeld type action.
BPS saturation is realized when the brane has the minimal volume
while keeping the appropriate winding numbers.
We apply the idea to two cases, type IIA superstring
compactified on $T^4$ and $K3$.
The result is consistent with the string duality, and
the expected spectrum for the BPS states is reproduced. 
\end{abstract}

\vfill
\noindent
\rule{7cm}{0.5pt}\par
\vskip 1mm
{\small
\noindent
\S \hskip 3mm JSPS fellow
}
\end{titlepage}

%%%%%%%%%%%%%%%%%%%%%%%%%%%%%%%%%%%%%%%%%%%%%%%%%%%%%%%%%%%%%%%%%%%
\section{Introduction}
One of the surprising discovery in the recent study
of the string duality\cite{HT}--\cite{6S1}
may be that
the excitations of  the fundamental strings
are identified with the solitonic spectrum under the 
duality transformations.  Such an identification is
quite remarkable since these string solitons are in general
higher dimensional objects (so called $p$-branes)
and their physical properties such as the quantization
are by far more complicated than strings.

There are some proposals to realize the solitons, namely,
the black-branes \cite{DGHR}
and more recently the Dirichlet (D-)branes \cite{DLP}--\cite{BVS2}.
The latter approach made it possible to use the technique of
the conformal field theory \cite{FT,CLNY,IO,Li1,CK}
to analyze the properties of solitons.
In particular, one may apply the results of
the open string theory to derive their effective
action.  It takes the form of the Born-Infeld action which
is similar to the Nambu-Goto action for $p=1$ case.
This effective action carries many essential properties of the
D-branes \cite{L,CK,Town,Sch}.  

In this letter, we derive some of the BPS masses of the D-branes
from the Born-Infeld action.
Together with the state counting \cite{Se2}--\cite{BVS2}, 
our result strongly supports string duality.
The examples that we study are
(i) U-duality \cite{HT,W2,SV}
in type II superstring compactified on 
the 4-dimensional torus, and (ii) duality between type IIA
theory compactified on K3 and heterotic string compactified
on the torus \cite{HT}--\cite{6S1}. These examples are known to have
massive BPS states in the fundamental string spectrum due to the
compactification. 
We show that this spectrum is successfully reproduced from the D-brane
calculation. 

%%%%%%%%%%%%%%%%%%%%%%%%%%%%%%%%%%%%%%%%%%%%%%%%%%%%%%%%%%%%%%%%%%%
\section{D-brane mass from Born-Infeld action}
Before we embark on the discussion in the specific models, let us
explain our strategy to study the D-brane mass.
We start from the Born-Infeld action for a Dirichlet $p$-brane
\cite{L,Sch},
\ba
\label{action}
S&=& \int {d^{p+1}}\sigma \, {e^{-{\phi}}} 
\sqrt{\det ({{G_{\mu\nu}}+{{\cal F}_{\mu\nu}}})} +
\int \exp (\frac{1}{2}{\cal F}) \wedge C \,\,\,,\\ 
{\cal F}&=& B + F\, ,
C = \sum_{i} C^{(i)} \, .\nonumber
\ea
Here $F$ is the gauge field on the world-volume and $C^{(i)}$ is an 
$i$-form arising from the Ramond-Ramond (R-R) sector ($i$ runs odd integer 
for type IIA theory and even integer for type IIB). 
Strictly speaking, since we are interested in   the supersymmetric 
D-branes, this effective  action \eqn{action} must include the fermionic part.
For our purpose, however, it is sufficient to work with only the bosonic 
sector.
 
Upon the string-string duality, the elementary excitations of 
strings with some Kaluza-Klein and winding charges should be 
mapped to the solitonic excitations with the corresponding 
R-R charges.  These objects  must appear as the
D-branes appropriately wrapping around a cycle in
the internal space $K$ ($T^4$ or $K3$ in this letter). 
From the six-dimensional point of view, this phenomenon is 
observed  as a correspondence between charged particles, 
especially, the BPS saturated particles, 
since the compactified space $K$ is invisible.

We calculate the effective 
mass of particles associated with D-branes in the 
following way. The relevant configuration of world brane 
is the form of ${{\br}\times \Sigma}$, where
${\br}$ gives the time axis and $\Sigma$ is 
a $p$-dimensional subspace of $K$.
For this configuration, we can rewrite 
the Born-Infeld action as
\beq
{\int_{\br}}ds\, e^{-\Phi} 
\left({{\int_\Sigma}{d^p}\sigma\, e^{\Phi - \phi}
\sqrt{\det (G+{\cal F})} }\right) = M_{p} \int_{\br}ds\, e^{-\Phi}\,\,\,.
\eeq
Here we introduced the six-dimensional dilaton
$\Phi = \phi - \frac{1}{4}\log \det G^{(K)}$ ($G^{(K)}$ is the 
metric of $K$). Clearly, the above equation 
represents a particle in six-dimension with mass $M_{p}$
\be
\label{D-mass}
  M_{p} = 
  {{\int_\Sigma}{d^p}\sigma\,e^{\Phi - \phi}\sqrt{\det (G+{\cal F})} }\,\,\,.
\ee
Roughly speaking, this is nothing but the volume of $\Sigma$ embedded in $K$.
It can be arbitrarily large if we change the embedding.  
However, for the BPS saturated state, we expect that
the configuration with minimal volume will be realized,
since the BPS state should be  the lightest state in each charged sector.
To be more precise, it is known \cite{BVS2} that the BPS condition 
is satisfied by the ``supersymmetric $p$-cycles'' 
\cite{BBS}. For a 2-brane, this is equivalent to 
a holomorphically embedded surface, which actually leads 
to the embedding with minimal area. It may be an interesting problem
whether the BPS condition leads to the minimal volume embeddings 
also in the general $p$-brane cases.

In the following sections, we calculate it
when the internal space $K$ is $T^4$ (type II U-duality) and 
$K3$ (heterotic-type II duality).

%%%%%%%%%%%%%%%%%%%%%%%%%%%%%%%%%%%%%%%%%%%%%%%%%%%%%%%%%%%%%%%%%%%
\section{BPS states in fundamental string spectrum}
The models we are considering as the
fundamental string are (i) Type II superstring and 
(ii) Heterotic string 
both  compactified on $T^4$.  In this section, we briefly review their 
moduli space and the mass spectrum
of the BPS states.

We first treat the type II string compactified on ${T^4}$.
The moduli space of this model is parametrized by the scalar fields in 
six dimension, from both of the NS-NS and the R-R sectors.
If we restrict ourselves to the case of vanishing R-R fields, the
moduli space is described by $G_{ij}$ and $B_{ij}$ on $T^4$ and takes
the form \cite{GPR}
\beq
O({4,4;{\bz}})\backslash O({4,4})/( O(4)\times O(4) )\,\,\,.
\eeq

In addition to the scalar fields, 
we have 16 gauge fields; 8 from the NS-NS sector and 8 from the R-R
one. 
The states in the elementary excitation are specified by 
the momenta $n_{i}$ and the winding numbers $m^{i}$
coupled with the NS-NS gauge fields.
The left- and the right-moving momenta 
${({p_L \,,\, p_R })}$ are defined as
\beq
p^a_R=({}^t\vec{n}+{}^t\vec{m}(B+G)){e^{\ast}}_a,
\qquad
p^a_L=({}^t\vec{n}+{}^t\vec{m}(B-G)){e^{\ast}}_a \, ,
\eeq
where ${{e^{\ast}}^i_a}$ is the dual vielbein 
${{\sum^{4}_{a=1}}{{e^{\ast}}^i_a}{{e^{\ast}}^j_a}={G^{ij}}}$.

The BPS states are states that break not all but a part of the space-time 
supersymmetry, and realized by setting the right-oscillator level 
$N_{R}=0$ (or the left-oscillator level $N_{L}=0$). 
The mass of the BPS states are given by 
$p_{R}^{2}$ (or $p_{L}^{2}$). More explicitly,
\beq
  {M^2}&=&{p_R^2} \nonumber \\
  &=& {}^t\vec{n}\,G^{-1}\vec{n}+
  2\,{}^t\vec{m} (1 - G^{-1}B)\vec{n}
  +{}^t\vec{m}(G-BG^{-1}B)\vec{m}\,\,\,.
\label{eqn:mass6}
\eeq
We will show that this spectrum is reproduced from the D-brane calculation. 

Next we turn to the case of the heterotic string on $T^{4}$.
This time, we have 24 massless gauge fields, and the momenta of the
physical states take the value in the Narain lattice
${{\Gamma}^{4,20}}$ \cite{Na1}.
The moduli space is parametrized by this lattice and written as
\beq
O({4,20;{\bz}})\backslash O({4,20})/( O(4)\times O(20) )\,\,\,. 
\eeq
The BPS states are again defined by the condition $N_{R}=0$
(Note that we can not adopt $N_{L}=0$,
since we have supersymmetry only on the right-moving sector),
and the mass is given by 
\be
\label{eqn:mass3}
M^2 = p_{R}^2 \, .
\ee

%%%%%%%%%%%%%%%%%%%%%%%%%%%%%%%%%%%%%%%%%%%%%%%%%%%%%%%%%%%%%%%%%%%
\section{U-duality in Type II theory and D-brane Mass Formula}
We first consider the case of the U-duality in type II theory 
compactified on $T^4$ \cite{SV,Se2,V2}. 
Under the U-duality transformation,
the BPS states of the fundamental string are mapped into the
$p$-brane BPS states with the R-R charges.
In the IIA (IIB) case, $p$ is an even (odd) integer.
The T-duality \cite{KY,GPR} on $T^4$ interchanges type IIA 
with IIB \cite{DLP,DHS}
since it flips the chirality. Without losing the generality,
we will examine the type IIA theory.

The moduli space of type II theory compactified on $T^4$ is
parametrized by the $T^4$ moduli and the scalar fields originated 
from R-R fields.
In this letter, we turn off the R-R fields and restrict ourselves to
the background with the diagonal metric and vanishing $B$ field
(This is because of the clarity of the discussion, and the 
following argument can be extended to the case of generic 
backgrounds \cite{IMSS}).
This type of background is characterized by the radii 
$(R_{6},R_{7},R_{8},R_{9})$ and the value of the dilaton $\phi$
(We take the coordinates of $T^4$ as $x^6, \cdots, x^9$). The BPS 
mass formula (\ref{eqn:mass6}) for this background takes the form 
\be
  \label{U-predicted}
  M^2 =  \sum_{i=1}^{4} 
  \left(\frac{1}{R_{i}} n_{i} + R_{i} m_{i} \right)^2 \, .
\ee

We pick a specific form of 
the U-duality transformation defined as
\beq
U=T^9 S\, T^{6789} S^{-1}\, T^{9}.
\eeq
Here $T^{i..j}$ means the T-duality transformation(s) in $x^i\cdots
x^j$ direction, while $S$ is the S-duality transformation in type IIB 
theory \cite{Schw1} that flips the sign
of the ten-dimensional dilaton $\phi$.
Under these transformations, the dilaton and the
moduli transform as \cite{GPR}
%%%%%%%  
\ba
 T^i : & &
 \begin{array}{rcl}
   \phi  &\longrightarrow & \phi - \log R_{i} \\
   R_{i} &\longrightarrow & 1/R_{i}
 \end{array} \\
   S : & &
 \begin{array}{rcl}
   \phi  &\longrightarrow & -\phi \\
   R_{i} &\longrightarrow & e^{-\phi/2} R_i \,\, .
 \end{array}
\ea
Here we used the fact that the Einstein metric
$g_{ij} = e^{-\phi/2} R_i^2 \delta_{ij}$ is invariant under the
S-duality transformation. For the transformation $U$, we combine
the above results to obtain
\ba
   \phi  &\longrightarrow & 
   \phi' = -\phi + \frac{1}{2}\log \frac{R_6 R_7 R_8}{R_9} \non
   R_{i} &\longrightarrow & (R_6',R_7',R_8',R_9') \\
   & & = \left(\sqrt{\frac{R_7 R_8}{R_6 R_9}},
         \sqrt{\frac{R_6 R_8}{R_7 R_9}},
         \sqrt{\frac{R_6 R_7}{R_8 R_9}},
         \sqrt{\frac{1}{R_6 R_7 R_8 R_9}}\right)\, . \nonumber
\ea

The transformation $U$ maps the NS-NS gauge fields to the R-R ones \cite{Se2}:
\be
\begin{array}{rclcrcl}
G_{6\mu}&\rightarrow& C_{69\mu},   & & B_{6\mu}&\rightarrow& C_{78\mu},\\
G_{7\mu}&\rightarrow& C_{79\mu},   & & B_{7\mu}&\rightarrow& C_{86\mu},\\
G_{8\mu}&\rightarrow& C_{89\mu},   & & B_{8\mu}&\rightarrow& C_{67\mu},\\
G_{9\mu}&\rightarrow& C_{6789\mu}, & & B_{9\mu}&\rightarrow& C_{\mu}.
\end{array}
\ee
Since the R-R fields $C_{\mu}, C_{ij\mu}$ and $C_{ijkl\mu}$ couple to 
0-, 2- and 4-branes respectively, the elementary excitation of string 
is mapped to a D-brane appropriately wrapped around the internal $T^4$. 
For example, the state with $n_{6} = N$ is realized by a 2-brane 
wrapping $N$ times around a 2-torus $T^{69}$. 

We can calculate the mass of these D-branes by using the formula 
(\ref{D-mass}) in the previous section. We first treat the case that 
only one kind of brane appears.

\

\noindent
\underline{\it 0-brane}

For 0-branes, the integral in the mass formula simply gives the number 
of branes $m_{9}$. Hence, the mass is written as 
\be
\label{0-mass}
  M_{0} =  e^{\Phi' - \phi'} m_{9} =  R_{9} m_{9} \, ,
\ee
where we used the value of the six-dimensional dilaton $\Phi'$ after the $U$ 
transformation
\be
  \Phi' = \phi' - \frac{1}{4}\log (R_{6}'R_{7}'R_{8}'R_{9}')^2
        = \phi' + \log R_{9} \, .
\ee
The mass derived above coincides with the value for the elementary 
spectrum (\ref{U-predicted}).

\

\noindent
\underline{\it 2-brane}

Let us parametrize a 2-brane $\Sigma$ by the 
coordinates $(s,t)$. 
The mass formula for $\Sigma$ reads
\be
  M_{2} =  e^{\Phi' - \phi'} \int_{\Sigma} ds dt 
  \sqrt{\det G^{(\Sigma)}}
  =  R_{9} \int_{\Sigma} ds dt \sqrt{\det G^{(\Sigma)}} \, ,
\ee
where $G^{(\Sigma)}$ is the induced metric of $\Sigma$. 
In order to evaluate this integral, 
we rewrite $\det G^{(\Sigma)}$ as follows
\be
  \det G^{(\Sigma)} = 
    (Q^{69} + Q^{78})^2 + (Q^{79} + Q^{86})^2 + (Q^{89} + Q^{67})^2 
    \, ,
\ee
where $Q^{ij}$ is defined as
\be
Q^{ij}= R_{i}' R_{j}' \left(
   \frac{\partial X^i}{\partial s}\frac{\partial X^j}{\partial t}
   -\frac{\partial X^i}{\partial t}\frac{\partial X^j}{\partial s} 
   \right).
\ee
The above integral is then bounded from below 
\beqs
  \lefteqn{
  \left(\int_{\Sigma} ds dt \sqrt{\det G^{(\Sigma)}}\right)^2 \geq
  }\\
  & &(R_{6}' R_{9}' N^{69} + R_{7}' R_{8}' N^{78})^2 +
     (R_{7}' R_{9}' N^{79} + R_{8}' R_{6}' N^{86})^2 +
     (R_{8}' R_{9}' N^{89} + R_{6}' R_{7}' N^{67})^2 \\
  & & = \frac{1}{R_{9}^2} \left(
     (\frac{1}{R_{6}} N^{69} + R_{6} N^{78})^2 +
     (\frac{1}{R_{7}} N^{79} + R_{7} N^{86})^2 +
     (\frac{1}{R_{8}} N^{89} + R_{8} N^{67})^2 \right) \, .
\eeqs
Here we used the Minkowski inequality
\be
\left(\int dsdt\left(\sum_{i=1}^3 f_i(s,t)^{2}\right)^{1/2}\right)^2 \geq
\sum_{i=1}^3 \left(\int |f_i(s,t)| ds dt \right)^2 \, ,
\label{eq:Minkowski}
\ee
and $N^{ij}$ is the winding number of $\Sigma$ about the $(i,j)$-torus.

As is noted before, this bound should be saturated for the BPS states.
Substituting the momenta $n_{i}$ and the winding number $m_{i}$ of the fundamental string in the above equation, we finally obtain the 
2-brane mass in the following form
\be
\label{2-mass}
 M_{2} =  \left(
     (\frac{1}{R_{6}} n_{6} + R_{6} m_{6})^2 +
     (\frac{1}{R_{7}} n_{7} + R_{7} m_{7})^2 +
     (\frac{1}{R_{8}} n_{8} + R_{8} m_{8})^2 \right)^{1/2} \, .
\ee
This is precisely the mass (\ref{U-predicted}) for the elementary 
spectrum. 

\

\noindent
\underline{\it 4-brane}

For a 4-brane, the integral in the mass formula (\ref{D-mass}) gives 
the winding number times the volume of the 4-torus.
Therefore, we get
\be
\label{4-mass}
  M_{4} =  R_{9} \times n_{9} \times R_{6}'R_{7}'R_{8}'R_{9}'
   = \frac{1}{R_{9}} n_{9} \, .
\ee
Again we reproduce the value of the elementary spectrum.

\ 

Thus we have shown that the mass of D-brane is consistent 
with the U-duality in type II theory.
In order to make this statement complete, we have to consider 
the case that the different kinds of branes coexist to form bound states
\cite{W1}. 
We will come to this point at the end of this letter.

%%%%%%%%%%%%%%%%%%%%%%%%%%%%%%%%%%%%%%%%%%%%%%%%%%%%%%%%%%%%%%%%%%%
\section{Heterotic-Type II duality}
The string duality of heterotic string on $T^4$ and type IIA
superstring on $K3$ \cite{HT}--\cite{6S1} is based on the isomorphism of their
moduli spaces \cite{Sei,As}. 
They share the same global structure,
\beq
O({4,20;{\bz}})\backslash
{O(4,20)}/({O(4)\times O(20)}) .
\eeq
This isomorphism  is based on the identification between the 
Narain lattice $\Gamma^{4,20}$ and
the integral cohomology ring  of $K3$ ($H^{*}(K3;{\bf Z})$).
Through this identification, the Kaluza-Klein
and  winding modes of the fundamental 
string spectrum are related to the 
winding numbers of the (Dirichlet) $p$-branes 
on the corresponding SUSY cycles.

%%%%%%%%
Let us first derive  the BPS mass formula for 
the D-brane that  is expected from the string duality.
We will later compare it with the mass formula from
the Born-Infeld action \eqn{D-mass}.
%%%%%%%%

For this purpose, we need specify the rule to translate 
the information of Narain lattice into those  of the cohomology of 
$K3$.
We introduce the intersection form
$ \dsp 
\alpha \cdot \beta := {\int_{K3}} \, 
\alpha\wedge\beta 
$
for
$
 \alpha,
\beta \in {H^{\ast}}(K3;{\br})
$
as an inner product on $H^{\ast}(K3;{\br})$. 
With this product, $H^{\ast}(K3;{\br})\cong {\br}^{4,20} $ holds,
and we must  pick up  the basis $\{ e_i \}_{i=1}^4$ of the positive norm
subspace (corresponding to the right moving momenta in the heterotic
side) in order to write down the desired mass formula.
The 2nd cohomology group $H^{2}(K3;{\bf Z})$ defines a lattice
$\Gamma^{3,19}$ from which we choose three orthonormal
basis with positive norm as $J_1,J_2,J_3$.
They are essentially three almost complex structures
on $K3$.
From $H^0$ and $H^4$ we
pick up the fourth vector.
We denote $u$ (resp. $v$) as a generator of
$H^0$ (resp. $H^4$) such that
$u\cdot v=1$. 
We express the two-form part of $e_4$ as $B$
which will be identified with the Kalb-Ramond field on $K3$.
According to the studies in  \cite{As},
we can exhibit the desired four orthonormal basis as
\beq
&&{e_i}={J_i}-({{J_i}\cdot B})v\,\,\,\,\,\,
(i=1,2,3)\,\,\,, \nonumber \\
&&\displaystyle {e_4}=\frac{1}{\sqrt{2}}\, \left\{ B
  +   ({1- \frac{1}{2}(B\cdot B)})
v+u    \right\} \, . \label{eqn:base2}
\eeq

Now, recalling   the mass formula for the fundamental string
\Eq{eqn:mass3}, we can present the mass formulae  for D-branes predicted 
by string duality.
Take   ${\br}\times \Sigma$ as the configuration of world brane.
Under the identification by Poincar\'{e} duality,
we obtain the following mass formula;  
\be
M^2 = \sum_{i=1}^4 (e_i\cdot \Sigma)^2.
\label{eq:mass7}
\ee

%%%%%%%%%%%%%%%%%%%%%%%%%%%%%%%%%%%%%%%%%%%%%%%%%%%%%%%%%%
We calculate the explicit forms of this formula
applied to 0-, 2-, 4-branes.

\ 

\noindent
\underline{\em{0-brane}} 

We write ${\alpha =n v}$, ${n \in {\bz}}$.
Eq.\Eq{eq:mass7} gives $M^2=\frac{1}{2}n^2$.
%%%%%%%%%%%%

~

%%%%%%%%%%%%
\noindent
\underline{\em{2-brane}} 

Let us write the embedded 2D surface as
$\Sigma$. Eq.\Eq{eq:mass7} simplifies to
$
M_2^2={\sum^{3}_{i=1}}
{{({{J_i}\cdot \Sigma})}^2}+\frac{1}{2}{{({{B}\cdot \Sigma})}^2}.
$
Later we consider the holomorphic embedding into $K3$.
In such case, it reduces to
\be
M_2^2=
\left(\int_\Sigma J\right)^2+ 
\frac{1}{2}\left(\int_\Sigma B\right)^2\, ,
\label{2brane-mass}
\ee
with the K\"ahler form $J$ on $K3$.
%%%%%%%%%%%%

~

%%%%%%%%%%%%
\noindent
\underline{\em{4-brane}}

The Poincar{\'e} dual of an arbitrary 
4-brane ${\alpha}$ is described
as
$
\alpha =n u
$
($n\in {\bz}$).
Plug it into \Eq{eq:mass7},
\be
M_4^2 =
{{n}^2}\left\{{
{{({{J_i}\cdot {B}})}^2}-\frac{1}{2} {{({{B}\cdot {B}})}}
 +\frac{1}{2}+\frac{1}{8} {{({{B}\cdot {B}})}^2} }\right\}\, .
\label{4brane-mass}
\ee
 
~

Next  we calculate the D-brane masses  by evaluating 
the Born-Infeld  actions and compare them with the above results. 
Evaluation of this integral is trivial
for the 0-brane case.

To give formulae for two-branes, which is holomorphically embedded in $K3$,
we first observe 
\be
\sqrt{\det (G+B)}=\sqrt{\det G + \det B}
\ee
for $2\times 2$ matrices.  
%As is already pointed out, 
%the SUSY 2-cycle
%around which the $D$ 2-brane should be wrapping is 
%a holomorphically embedded surface.  
In order to get a concrete answer, we would postulate that we can choose
those matrices as constant.
It may be justified  in the orbifold limit
of $K3$.
Under this assumption, we may identify 
$
\dsp  \det G = 2 \, \left( \int_\Sigma J\right)^2 
$ and
$\dsp 
\det B= \left( \int_\Sigma B \right)^2
$.
Hence the mass formula from the Born-Infeld action \eqn{D-mass} is 
reduced to the expected formula \Eq{2brane-mass}.

Similar reasoning is also applicable to  the 4-brane case.
We first observe the following  identity
\beq
\det (G+B)=\left({
1+\frac{1}{2}{G^{ac}}{G^{bd}}{B_{ab}}{B_{cd}} }\right)
\det G +\det B ,
\eeq
where $G^{ab}$ means the components of the inverse matrix 
of $G$.
Under the same assumption as above, the right hand side is equal to
\be
1+  \int B\wedge \ast B+ \frac{1}{4}
{{({B\cdot B})}^2}   \,\,\,.
\ee
After some computation this leads to the expected 
result \Eq{4brane-mass}.

%%%%%%%%%%%%%%%%%%%%%%%%%%%%%%%%%%%%%%%%%%%%%%%%%%%%%%%%%%%%%%%%%%%
\section{Discussion}
In this letter, we have shown that the BPS mass formula for the 
elementary excitations can be reproduced from the D-brane calculation, 
which provides us with an evidence that D-branes are the very objects 
required by string duality. 

In order to complete the discussion, we have to treat the case
that different kinds of branes coexist.

We first describe the bound states of $0$- and $2$-branes. This state 
can be realized by giving non-vanishing value to the field strength 
$F$ on the 2-brane world-volume
\cite{W1,D,V1}. Suppose a 2-brane $\Sigma$ 
parametrized by $(s,t)$ and the field strength $F$ in the form of 
$2 N ds dt$. Then, the Chern-Simons term in the D-brane action 
(\ref{action}) reads
\be
  \int_{\Sigma\times\br}e^{\frac{1}{2}F} \wedge C = 
  \int_{\Sigma\times\br} C^{(3)} + N \int_{\br} C^{(1)} \, .
\ee
Therefore, this 2-brane has $N$ units of 0-brane charge besides its 
coupling to the 3-form $C^{(3)}$ and is considered to be a bound state 
of a 2-brane $\Sigma$ with $N$ 0-branes. 

The mass of this state is easily calculated by using our formula 
(\ref{D-mass}). From the Minkowski inequality (\ref{eq:Minkowski}), the 
integral in the formula is evaluated as
\be
  \left(\int_{\Sigma} ds dt \sqrt{\det(G + F)} \right)^{2} \geq
  \left(\int_{\Sigma} ds dt \sqrt{\det G} \right)^{2} + N^{2} \, .
\ee
Together with the dilaton factor, we finally obtain the mass of the bound 
state in the following form
\be
  M^{2} = (M_{2})^{2} + N^{2} (M_{0})^{2} \, .
\ee
This is exactly the value of the elementary BPS spectrum. 

For the case of 2- and 4-branes, we can proceed in the same way as above. In this case, non-vanishing field strength gives 2-brane 
charge to the 4-brane. One can show that application of our mass 
formula again yields the correct result. 
However, there is one subtlety. 
For a generic configuration, $F \wedge F$ does not vanish  
and the 4-brane has both of 0- and 2-brane charges, which corresponds 
to a bound state of 0-, 2- and 4-branes. 
The problem is that the 0-brane charge $F \wedge F$ is 
not arbitrary and depends on
the 2-brane charge $F$.
This is not a satisfactory situation,
since 0-, 2- and 4-brane charges are mutually independent quantity, 
say, windings and momenta, on the side of the elementary excitation. 
One way to avoid this difficulty may be to consider a non-abelian gauge 
field, for which tr$F\wedge F$ can be independent of tr$F$. The 
appearance of non-abelian gauge fields seems to be natural, since a 
bound state is composed of multiple branes for which we can expect 
non-abelian gauge groups. We think this point deserves further 
investigation. 

\ 

This work is supported in part by the Grant-in-Aid for Scientific
Research from the Ministry of Education, Science and Culture.
The work of H.I. and K.S. is supported by the Japan Society for the
Promotion of Science, and Y.S. is supported by Soryushi-Shogaku-Kai.

%%%%%%%%%%%%%%%%%%%%%%%%%%%%%%%%%%%%%%%%%%%%%%%%%%%%%%%%%%%%%%%%%%%


\newpage

\baselineskip=0.55cm
\begin{thebibliography}{99}
%%%%%%%%%%%%%%%%%%%%%%%%
%  String Duality   %%%%%%%%%
%%%%%%%%%%%%%%%%%%%%%%%%%%%%
\bibitem{HT} C.~Hull and P.~Townsend, 
%``Unity of Superstring
%Dualities'', QMW-94-30, R/94/33, {\hp}9410167, 
Nucl.~Phys. {\bf B438} (1995) 109, {\hp}9410167.
\bibitem{W2} E.~Witten, %``String Theory Dynamics in Various
%Dimensions'', 
Nucl.~Phys. {\bf B443} (1995) 85, {\hp}9503124.
%%%%%%%%%%%%%%%%%%%%%%%%%%
%  6-dim duality   %%%%%%
%%%%%%%%%%%%%%%%%%%%%%%%%%
\bibitem{HS} J.~Harvey and A.~Strominger,
%``The Heterotic String is a Soliton'', hep-th/9504047,
Nucl.~Phys. {\bf B449} (1995) 535, hep-th/9504047.
\bibitem{6S1} A.~Sen, %``String String Duality Conjecture in Six
%Dimensions and charged Solitonic Strings'', TIFR-TH-95-16,
Nucl.~Phys. {\bf B450} (1995) 103, {\hp}9504027.
%\bibitem{6S2} A.~Sen, ``Strong-Weak Coupling Duality in Four
%Dimensional String Theory'', TIFR/TH/94-03, 
%Int.~J.~Mod.~Phys. {\bf A9} (1994) 3707, {\hp}9402002;
%``Strong-Weak Coupling Duality in Three Dimensional String Theory'', 
%Nucl.~Phys.~{\bf B434} (1995) 179, {\hp}9408083.
%%%%%%%%%%%%%%%%%%%%%
%  Black Brane   %%%%%%%%%
%%%%%%%%%%%%%%%%%%%%%%%%%%
\bibitem{DGHR} A.~Dabholkar, G.~Gibbons, J.~A.~Harvey and
F.~Ruiz~Ruiz, %``Superstrings and Solitons'', 
Nucl.~Phys. {\bf B340} (1990) 33.\\
%\bibitem{HS} 
G.~Horowitz and A.~Strominger, %``Black Strings and ${p}$-Branes'', 
Nucl.~Phys. {\bf B360} (1991) 197.\\
%\bibitem{St} 
A.~Strominger, 
%``Massless Black Holes and Conifolds in String Theory'', {\hp}9504090,
Nucl.~Phys. {\bf B451} (1995) 96, {\hp}9504090.
%%%%%%%%%%%%%%%%%%%%%%%%%%%%
% Born-Infeld action   %%%%%%%
%%%%%%%%%%%%%%%%%%%%%%%%%%%%%
\bibitem{DLP} J.~Dai, R.~G.~Leigh and J.~Polchinski, 
%``New Connections between String Theories'',
Mod.~Phys.~Lett. {\bf A4}~(1989) 2073.
\bibitem{L} R.~G.~Leigh, 
%``Dirac-Born-Infeld Action From Dirichlet ${\sigma}$-Model'',
Mod.~Phys.~Lett. {\bf A4}~(1989) 2767.
\bibitem{P1} J.~Polchinski, 
%``Combinatorics of Boundaries in String Theory'', NSF-ITP-94-73, 
Phys. Rev. {\bf D50} (1994) 6041, {\hp}9407031.
\bibitem{P2} J.~Polchinski, 
%``Dirichlet-branes and Ramond-Ramond charges'', NSF-ITP-95-122, 
Phys. Rev. Lett. {\bf 75} (1995) 4724, {\hp}9510017.\\
%\bibitem{P3} 
J.~Polchinski, S.~Chaudhuri and C.~Johnson, 
``Notes on D-Branes'', NSF-ITP-96-003, {\hp}9602052.
\bibitem{W1} E.~Witten, %``Bound States of Strings and ${p}$-Branes'', 
%IASSNS-HEP-95-83, 
Nucl.~Phys. {\bf B460} (1996) 335, {\hp}9510135.
\bibitem{Li1} M.~Li, Nucl.~Phys. {\bf B460} (1996) 351, {\hp}9510161.
\bibitem{W3} E.~Witten, %``Small Instantons in String Theory'',
%IASSNS-HEP-95-87, 
Nucl.~Phys. {\bf B460} (1996) 541, {\hp}9511030.
\bibitem{CK} C.~G.~Callan and I.~R.~Klebanov, ``D-Brane Boundary State
Dynamics'', PUPT-1578, {\hp}9511173.
%%%%%%%%%%%%%%%%%%%%%%
\bibitem{Town} P.~K.~Townsend, ``D-Branes from M-Branes'',
DAMTP-R-95-59, {\hp}9512062.
\bibitem{D} M.~Douglas, ``Branes within Branes'', RU-95-92, {\hp}9512077.
\bibitem{Sch} C.~Schmidhuber, ``D-brane actions'', PUPT-1585, {\hp}9601003.
%%%%%%%%%%%%%%%%%%%%%%%%%%%%%%%%%%%%%%
\bibitem{Se2} A.~Sen, ``A Note on Marginally Stable Bound States in
Type II String Theory'', MRI-PHY/23/95, {\hp}9510229.\\
%\bibitem{Se1} 
A.~Sen, %``U-Duality and Intersecting D-Branes'', 
%MRI-PHY/27/95
Phys.~Rev. {\bf D53} (1996) 2874, {\hp}9511026.
%%%%%%%%%%%%%%%%%%%%%%%%%%%%%%%%%
%  degeneracy counting   %%%%%%%%
%%%%%%%%%%%%%%%%%%%%%%%%%%%%%%%%%%
\bibitem{V2} C.~Vafa, ``Gas of D-Branes and Hagedorn Density of BPS
States'', HUTP-95/A042, {\hp}9511088.
\bibitem{V1} C.~Vafa, ``Instantons on D-branes'', HUTP-95/A049, {\hp}9512078.
%\bibitem{BVS1} M.~Bershadsky, C.~Vafa and V.~Sadov, 
%``D-Strings and D-manifolds'', IASSNS-HEP-95-77, HUTP-95/A035, {\hp}9510225.
\bibitem{BVS2} M.~Bershadsky, C.~Vafa and V.~Sadov, 
``D-Branes and Topological Field Theories'', HUTP-95/A047, {\hp}9511222.\\
%\bibitem{YZ} 
S.-T.~Yau and E.~Zaslow, ``BPS States, String Duality,
and Nodal Curves on K3'', {\hp}9512121.

%%%%%%%%%%%%%%%%%%%%%%%%%
%   CFT  %%%%%%%%%%%%%%
%%%%%%%%%%%%%%%%%%%%%
\bibitem{FT} E.~Fradkin and A.~Tseytlin, 
Phys.~Lett. {\bf B163} (1985) 123.
\bibitem{CLNY} C.~G.~Callan, C.~Lovelace, C.~R.~Nappi and S.~A.~Yost, 
Nucl.~Phys.~{\bf B288} (1987) 525, 
Nucl.~Phys.~{\bf B293} (1987) 83, 
Nucl.~Phys.~{\bf B308} (1988) 221. 
\bibitem{IO} N.~Ishibashi and T.~Onogi, 
%``Open String Model Building'', 
Nucl.~Phys.~{\bf 318} (1989) 239.

%%%%%%%%%%%%%%%%%%%%%%%%%%%%%%%%%%%%%%%%%%%%%%%
%  U-duality   %%%%%%
%%%%%%%%%%%%%%%%%%%%%%
\bibitem{SV} A.~Sen and C.~Vafa, %``Dual Pairs of Type II String
%Compactification'', HUTP-95/A028, TIFR/TH/95-41, {\hp}9508064,
Nucl.~Phys. {\bf B455} (1995) 165, {\hp}9508064.
%%%%%%%%%%%%%%%%%%%%%%
\bibitem{BBS} K.~Becker, M.~Becker and A.~Strominger, %``Fivebranes,
%Membranes and Non-perturbative String Theory'', NSF-ITP-95-62,
Nucl.~Phys. {\bf B456} (1995) 130, {\hp}9507158, .
%%%%%%%%%%%%%%%%%%%%%%%
%%%%%%%%%%%%%%%%%%%%%
%              %%%%%%%%%%%%
%%%%%%%%%%%%%%%%%%%%%%
%%%  T-dual    %%%%%%%%%%%
%%%%%%%%%%%%%%%%%%%%%%%%
\bibitem{GPR} A.~Giveon, M.~Porrati and E.~Rabinovici, 
Phys.~Rep.~{\bf 244} (1994) 77, and references therein.
\bibitem{Na1} K.~Narain, %``New Heterotic String Theories in
%Uncompactified Dimensions ${<1}$'', 
Phys.~Lett. {\bf B169} (1986) 41.\\
%\bibitem{Na2} 
K.~Narain, M.~Sarmadi and E.~Witten, %``A Note on Toroidal
%Compactification on Heterotic String Theory'', 
Nucl.~Phys. {\bf B279} (1987) 369.
\bibitem{KY} K.~Kikkawa and M.~Yamasaki, 
Phys.~Lett. {\bf B149} (1984) 357.\\
%\bibitem{Sa} 
N.~Sakai and I.~Senda, 
Prog.~Theor.~Phys. {\bf 75} (1986) 692.
\bibitem{DHS} M.~Dine, P.~Huet and N.~Seiberg,
%``Large and Small Radius in String Theory'', 
Nucl.~Phys.~{\bf B322} (1989) 301.
%%%%%%%%%%%%%%%%%%%%%%%%%
%%%%  S-Dual %%%%%%%%
%%%%%%%%%%%%%%%%%%%%%
\bibitem{Schw1} J.~H.~Schwarz, %``An SL({${2,{\bz}}$}) Multiplet of
%Type IIB Superstrings'', CALT-68-2013, 
Phys.~Lett. {\bf B360} (1995) 13, 
Erratum Phys.~Lett. {\bf B364} (1995) 252, {\hp}9508143.\\
%\bibitem{Schw2} 
J.~H.~Schwarz, %``The Power of M Theory'', RU-95-68,
Phys.~Lett. {\bf B367} (1996) 97, {\hp}9510086.
%%%%%%%%%%%%%%%%%%%%%%%%%%
%  susy cycle   %%%%%%%%%%
%%%%%%%%%%%%%%%%%%%%%%%%%
%\bibitem{BJSV} M.~Bershadsky, A.~Johansen, V.~Sadov and C.~Vafa, 
%``Topological Reduction of 4-D SYM to 2-D Sigma Models'',
%HUTP-95-A004, {\hp}9501096.
%%%%%%%%%%%%%%%%%%%
%  K3      %%%%%%%%
%%%%%%%%%%%%%%%%%%
\bibitem{Sei} N.~Seiberg, 
%``Observations on the Moduli Space of Superconformal Field Theories'',
Nucl.~Phys.~{\bf B303} (1988) 286.
\bibitem{As} P.~S.~Aspinwall and D.~R.~Morrison, ``String Theory on K3
Surfaces'', IASSNS-HEP-94/23, DUK-TH-94-68, {\hp}9404151.\\
%\bibitem{AM} 
P.~S.~Aspinwall, 
%``Enhanced Gauge Symmetries and K3 Surfaces'',
Phys.~Lett. {\bf B357} (1995) 329, {\hp}9507012.
%%%%%%%%%%%%%%%%%%%%%%
\bibitem{IMSS} H.~Ishikawa, Y.~Matsuo, Y.~Sugawara and K.~Sugiyama, 
in preparation.
%%%%%%%%%%%%%%%%%%%%%%
\end{thebibliography}
\end{document}